\documentstyle[twocolumn,prl,aps,epsf]{revtex}
\newcommand{\non}{\nonumber}

\newcommand{\ii}{{i}}
\newcommand{\be}{\begin{eqnarray}}
\newcommand{\en}{\end{eqnarray}}

\title{Staggered local
density-of-states around the vortex
in underdoped cuprates}

\author{Jun-ichiro Kishine\cite{JK-address},
Patrick A. Lee, and Xiao-Gang Wen}

\address{Department of Physics, 
Massachusetts Institute of Technology,
Cambridge, Massachusetts 02139}

\begin{document}

\twocolumn[\hsize\textwidth\columnwidth\hsize\csname @twocolumnfalse\endcsname

\maketitle

\begin{abstract}
We have studied 
a single vortex with the staggered flux (SF) core based on the SU(2) 
slave-boson theory of high $T_c$ superconductors. 
We find that whereas the center in the vortex core is a SF state, 
as one moves away from the core center, 
a correlated staggered modulation of the hopping amplitude $\chi$
and pairing amplitude $\Delta$  becomes predominant.
We predict that in this region,
the local density-of-states (LDOS) exhibits staggered modulation when measured 
on the bonds, which may be
directly detected by STM experiments.
\end{abstract}

\pacs{#1}

]

\narrowtext



The elucidation of the ground state properties of the high $T_c$  cuprates has
been one of the major challenges in  condensed matter physics.  Recent STM
experiments on vortices\cite{Renner98,Pan00} have indicated the existence of a
normal core with a gap structure characteristic of the normal state pseudo gap
above $T_c$.  The theoretical description of the normal core, however,  remains
unresolved\cite{Nagaosa92,Sachdev92,FT00}.  Based on the SU(2) slave-boson
theory of the high $T_c$ superconductors\cite{WenLee96,LNNW98},  Lee and
Wen\cite{LeeWen00} proposed a model of the vortex with a staggered flux (SF)
core, characterized by a pseudo gap and  staggered orbital current.  The signature
of the staggered flux (SF) order in zero field was found in  the
Gutzwiller-projected $d$-wave superconducting state by  variational Monte
Carlo method\cite{Ivanov00} and exact diagonalization of the $t$-$J$
model\cite{Leung00}.  By using a Gutzwiller-projected U(1) slave-boson
mean-field wave function, Han, Wang, and Lee recently found evidence of the SF
order near the vortex core\cite{Han-Wang-Lee,Han-Wang-Lee2}.  The natural
question arises as to whether the SF core has  observable consequences for the
tunneling spectra using atomic resolution STM. This is the question addressed
by the present paper.

We first summarize some of the salient feature of the SU(2) vortex structure.
In the SU(2) slave-boson theory\cite{WenLee96},
the physical electron is represented by a
spin-1/2 fermion operator $ f_{i\sigma}$ and a charge-1 boson operator
$h_{i}^T=(b_{i1},b_{i2})$:
$
c_{i\sigma}={1\over\sqrt{2}}h_{i}^{\dagger}\psi_{i\sigma},
$
where 
$ h_{i} $
and
$
\psi_{i\sigma}^T=(f_{i\sigma},
\varepsilon_{\sigma\bar\sigma}
f_{i\bar\sigma}^\dagger)
$
are SU(2) doublets.
The slave-boson mean-field state is characterized 
by $2\times 2$ matrices
$U_{ij}=\pmatrix{-\chi_{ij}^\ast&\Delta_{ij}\cr
 \Delta_{ij}^\ast&\chi_{ij}}$ on the links and 
boson condensation $h_i$ on the sites.
The $U_{ij}$ describe the hopping of the 
fermions and bosons. 
The mean-field solution 
is obtained by 
integrating out the fermions and
minimizing the mean-field energy $E(\{ U_{ij}, h_i\})$,
which leads to
$\chi_{ij}=\langle f_{i\sigma}^\dagger f_{i\sigma}\rangle$
and $\Delta_{ij}=\langle \varepsilon_{\sigma\bar\sigma} f_{i\sigma} 
f_{j\bar\sigma}\rangle$.
%
The SU(2) gauge invariance is realized through the relation
$ E(\{ U_{ij}, h_i\}) = E(\{ W_iU_{ij}W_j^\dagger, W_i h_i\})$, for any
$W_i \in {\rm SU(2)}$.
In the underdoped region, the mean-field solution has the following form
(in the staggered flux (SF) gauge)
$
U_{ij}^{\rm SF}
=-A
\tau^3
\exp[\ii(-1)^{i_x+j_y}\Phi_0\tau^3]
$
with 
$A=\sqrt{\chi^2+\Delta^2}$ and
$\Phi_0=\tan^{-1}(\Delta/\chi)$.
This describes fermions hopping with flux $\pm 4
\Phi_0$ on alternating plaquettes\cite{Affleck88}.
The advantage of the SF gauge is that 
it is apparent that the SU(2) symmetry 
has been broken down to the residual 
U(1), since $U_{ij}^{\rm SF}$ contains only $\tau^3$. 
As a result, the low lying fluctuations 
have a simple form in the SF gauge:
\begin{equation}
\label{Uija3}
\bar U_{ij}^{\rm SF}
=-A\tau^3
\exp\left[
i (-1)^{i_x+j_y}\Phi_0\tau^3\right]
\exp\left[-i a_{ij}^3\tau^3\right].
\end{equation}
These fluctuations are described by the
lattice gauge field $a_{ij}^3$. 

In the presence of a magnetic field, the mean-field solution contains vortices.
Since the bosons are locally condensed to the bottom of the band, they 
can be parameterized by
$
(\bar h_{i}^{\rm SF})^T
=\sqrt{x}(z_{i1}, -i(-1)^{i_x+i_y} z_{i2}),
$
where $z_{i1}$ and $z_{i2}$ are slowly varying in space and time.
The vortex is now described by
the twisted $z_i$ field and $U_{ij}$ in Eq.~(\ref{Uija3}).
To write down the vortex structure explicitly,
let us introduce
$
z_{i1}
=e^{\ii\varphi_{i1}}\cos{\theta_i\over 2}
$
and
$
z_{i2}=e^{\ii\varphi_{i2}}\sin{\theta_i\over 2},
$
where 
$
\varphi_{i1}
=\alpha_i-\phi_i/2$,
and
$
\varphi_{i2}=\alpha_i+\phi_i/2.
$
The phase angles $\alpha_i$ and $\phi_i$  are associated with
the electromagnetic (EM) U(1) and the $a^3$ U(1) gauge structures, 
respectively.
The internal degrees of freedom $\phi$ and $\theta$ can be
 visualized by the vector
$
{\bbox I}_i
=z_i^\dagger{\bbox\tau} z_i
=
(\sin\theta_i\sin\phi_i
,-\sin\theta_i\cos\phi_i,\cos\theta_i),
$
which has the meaning of the quantization axis for the $z_i$ fields.

In the 
vortex structure proposed by Lee and Wen\cite{LeeWen00},
both $\alpha_i$ and $\phi_i/2$ wind by
$\pi$ and consequently give
an appropriate $hc/2e$ vortex for
the EM 
gauge field ${\bbox A}({\bbox r})$, i.e.
$
{\bbox \nabla} \alpha={\bbox \nabla}
{ \phi/ 2}=\hat{\bbox e}_\phi/2r
$
which leads to
$
{\bbox \nabla} \varphi_1=0
$
and 
$
{\bbox \nabla} \varphi_2={\hat{\bbox e}_\phi
}/r
$, 
where $\hat{\bbox e}_\phi$ denotes the azimuthal unit vector in the physical 
space.
That is to say, only
$b_2$ changes its phase
$\varphi_2$ 
by $2\pi$ as we go around the vortex,
while 
$b_1$ does not.
Outside the vortex core, 
${\bbox I}_i
=(\sin\phi_i
,-\cos\phi_i,0)$    
(see Fig.~1(a)), which describes a SC state.
As we approach the core,
$|b_2|$ must vanish and the vortex center 
is represented by ${\bbox I}_i=(0,0,1)$,  which is just the metallic SF state.
The ${\bbox I}_i$-vector tilts smoothly from the equator to
the north pole as the core is approached
with a length scale denoted by $\ell_c$.
At the same time, the gauge flux 
${\bbox{h}}={\bbox\nabla}\times{\bbox a}^3$ is
distributed over a distance scale of $\lambda_a$ and it is expected
that 
$\lambda_a\sim x^{-1}
\agt\ell_c$.
Inside $\lambda_a$, the gauge invariant combination
\be
{\cal V}_{ij}=
{\phi_i-\phi_j\over2}-a_{ij}^3
\en
and its continuum limit,
$
{\bbox {\cal V}}({\bbox r})=
{1\over 2}{\bbox\nabla}\phi(\bbox{r})
-{\bbox a}^3({\bbox r}),
$
is finite. This is the analog of the superfluid velocity inside the
London penetration depth $\lambda_L$ in conventional vortices,
except that here $\lambda_{a}\ll\lambda_L$
 and the effect of the EM field
${\bbox A}$ is negligible.

The physics of the local electronic state is better visualized by making a 
local gauge transformation
$\psi_{i\sigma}\to g_i \psi_{i\sigma}$,
$\bar U_{ij}^{\rm SF}\to g_{i}
\bar U_{ij}^{\rm SF} g_j^\dagger$, and
$
(\bar h_{i}^{\rm SF})^T
\to (g_i \bar h_{i}^{\rm SF})^T
=e^{i\alpha_i}(\sqrt{x},0) 
$, i.e.\
the local boson isospin vector points toward the north pole, as shown in Fig.~1(b).
The explicit $g_i$ is given by
$
g_i=\exp[{\ii(-1)^{i_x+i_y}{\theta_i\over 2}\tau^1}]
\exp[{\ii{\phi_i\over 2}\tau^3}]
$.
The advantage of this gauge is that the physical electron operator
$c_\sigma \propto f_\sigma$.
The new $U_{ij}$
has a physical meaning as governing the hopping and pairing of electrons.
We shall refer to this as the $d$-wave gauge.
Indeed, locally there is a single boson component and the problem reduces to
the more familiar U(1) mean-field theory, but with $\chi_{ij}$ and 
$\Delta_{ij}$
which vary in space. This is precisely the problem treated by Han, Wang, and 
Lee\cite{Han-Wang-Lee,Han-Wang-Lee2} and it is gratifying that they found 
numerically the  staggered current around the vortex core as
proposed in Ref.\cite{LeeWen00}.

\begin{figure}
\epsfxsize=2.2in
\centerline{\epsffile{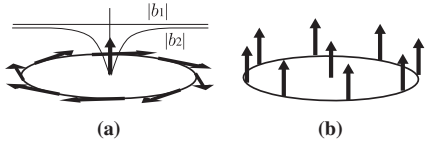}}
\smallskip
\caption{
(a) Configuration of the ${\bbox I}_i$-vector
 outside the SF core in the SF gauge.
At the center of the vortex,
${\bbox I}_i$ points toward the north pole, which corresponds to the  SF state.
The local gauge  transformation $g_i$ transforms this configuration
to (b) in the $d$-wave gauge, where the internal phases of the bose condensate 
are gauged away.
}
\end{figure}

Now we are ready to ask
the question:
is  any signature of the unit cell doubling  directly measurable by STM
tunneling?
It turns out that there is no effect in the SF phase in the center of the 
core, because
the period doubling of the {\it current} does not show up in the LDOS.
This leads us to look outside the immediate core and examine the effect of the 
phase
winding, i.e.\ the effect of ${\cal V}_{ij}$.
After a local gauge transformation to the $d$-wave gauge, we find 
Eq.~(\ref{Uija3}) becomes
\be
&&\bar U_{ij}^d
=-\chi_{ij} 
\left(\tau^3 \cos{\theta_i-\theta_j\over 2}
+\tau^2 \sin{\theta_i-\theta_j\over 2}
\right)\label{Uij}\\
&-&\Delta_{ij}\left[
\ii(-1)^{i_x+j_y}
\cos{\theta_i+\theta_j\over 2}
-(-1)^{i_y+j_y}\tau^1
\sin{\theta_i+\theta_j\over 2}\right]\non,
\en
where
\be
\chi_{ij}= A\cos\Phi_{ij},\,\,\,\,\,\,\,\,
\Delta_{ij}=A\sin\Phi_{ij},\label{modchiDelta}
\en
and
\be
\Phi_{ij}=\Phi_0+
{(-1)^{i_x+j_y}}
{\cal V}_{ij}.\label{Phiij}
\en

We now interpret this equation in different limits.
First, we consider the region far outside the SF core.
In this region,
$\theta_i\sim\theta_j\sim\pi/2$  and 
Eq.~(\ref{Uij}) becomes
$\bar U_{ij}^d\sim-\chi_{ij} \tau^3+(-1)^{i_y+j_y}\Delta_{ij} \tau^1$. 
Recalling that $\chi_{ij}$ and $\Delta_{ij}$ are interpreted as the
hopping and pairing amplitudes of physical electrons, we see that
{\it the region 
outside the SF core 
 is characterized by 
the staggered amplitude modulation}.
Note
from Eq.~(\ref{modchiDelta}) 
that the amplitude of $\chi_{ij}$ and
$\Delta_{ij}$ are modulated in a correlated way to preserve
$\chi_{ij}^2+\Delta_{ij}^2={\rm constant}$.
In Fig.~2, we schematically show the modulation pattern of 
$\chi_{ij}$.
In this region,
$\bar U_{ij}^d$ breaks
 the translational symmetry, but  does not break
 the time reversal symmetry. 

\begin{figure}
\epsfxsize=2.6in
\centerline{\epsffile{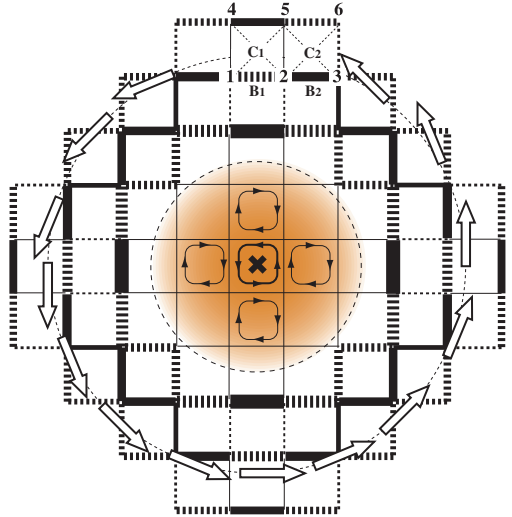}}
\smallskip
\caption{
Schematic drawing of 
the amplitude modulation pattern of the hopping  field
$\chi_{ij}$ outside the SF core.
Solid and dotted bonds indicate 
 enhanced and reduced amplitudes, respectively, where
thickness of the bonds qualitatively represents magnitude of the modulation. 
Circulation of the internal gauge field, ${\bbox {\cal V}}({\bbox r})$, 
 is  indicated by the arrows. 
The dotted circle represents the SF core boundary, $r=\ell_c$, inside which the 
staggered orbital
current flows. Note 
that the boundary should not be taken literally. In reality, 
there is a crossover region around
$r\sim \ell_c$ where the staggered 
current and the staggered amplitude modulation coexist. 
}
\end{figure}

Next, we consider the vicinity of the  vortex center,
where 
$\theta_i\sim\theta_j\sim 0$
 and Eq.~(\ref{Uij}) becomes
$\bar U_{ij}^d\sim
-A\tau^3
\exp\left[
i (-1)^{i_x+j_y}\Phi_{ij}
\tau^3\right]
$. 
In this region,  $\bar U_{ij}^d$ breaks
 both the translational and
 the time reversal symmetries, and the staggered orbital 
 current appears\cite{LeeWen00}.
However, the staggered flux is further modulated according to
Eq.~(\ref{Phiij}).

As we approach the core from the outside,
the ${\bbox I}_i$-vector in the SF gauge representation gradually rises off 
from the equatorial plane [see Fig.~1(a)]. 
This   gives rise to a
 crossover region characterized by
 coexistence of the amplitude and phase modulation,
 where  the $\theta$-dependence of $\bar U_{ij}^d$ 
 becomes significant.
However,  to study the effects of a $\theta$-dependent $\bar U_{ij}^d$ 
is beyond the scope of the present paper.
From now on,  we shall 
compute the LDOS by setting $\theta_i=\theta_j=\pi/2$.
We expect our results to be qualitatively valid for 
$r\agt \ell_c$
as long as we avoid the exact core center.

The presence of staggered modulation of 
$\chi_{ij}$ and $\Delta_{ij}$ 
in this region
 suggests that this
 may be the best place
 to look for unit cell doubling effects.
To model the tunneling current, we assume that the electrons tunnel from the
tip, located at 
 $\bbox r$, to a linear combination of Wannier orbitals centered at lattice 
sites $i$,
i.e. $\sum_{i}\alpha_i({\bbox r})\phi({\bbox r}-{\bbox r}_i)$.
Then the LDOS  in the $d$-wave gauge is written as
\be
N({\bbox r},\omega)=
-{2x\over \pi}{\rm Im}
\sum_{i,j}\alpha_i({\bbox r})\alpha_j({\bbox r})
[{\cal G}_{ij}^{\rm F}(\ii\omega)]_{11}
\mid_{\ii\omega=\omega+\ii\delta}.
\en
The subscript 11 means the 11-component of the
lattice fermion propagator in a $2\times 2$ matrix form,
$
{\cal G}_{ij}^{\rm F}(i\omega)=
-\int_0^\beta
 d\tau e^{i\omega\tau}
\langle T_{\tau} \psi_{i\sigma}(\tau)\psi_{j\sigma}^\dagger\rangle
$.

Here we  demonstrate that
the LDOS  exhibits conspicuous staggered pattern 
only when measured on the bonds.
For example, 
we pick up the sites $1,2,...,6$ indicated in Fig.~2
and consider the midpoints on the bonds, B$_1$, B$_2$, 
and the plaquette centers, C$_1$, C$_2$.
The LDOSs  at C$_1$ and C$_2$ come from
 $
\sum_{i,j=1,2,4,5}
{\cal G}_{ij}^{\rm F}$
and  $
\sum_{i,j=2,3,5,6}
{\cal G}_{ij}^{\rm F}$,
respectively. We see, however, that 
${\cal G}_{12}^{\rm F}\sim{\cal G}_{56}^{\rm F}$ 
since the bonds 12 and 56 are
almost equivalent except for a small inequivalence coming from the nonuniformity of 
the slowly-varying ${\bbox {\cal V}}({\bbox r})$ field. Similarly, 
${\cal G}_{45}^{\rm F}\sim{\cal G}_{23}^{\rm F}$ and
${\cal G}_{14}^{\rm F}\sim{\cal G}_{36}^{\rm F}$.
Therefore,  $N({\rm C}_1,\omega)\sim N({\rm C}_2,\omega)$.
Similarly, the LDOS
at the lattice sites is almost uniform.
On the other hand, the LDOSs 
at B$_1$ and B$_2$ come from
 $
\sum_{i,j=1,2}
{\cal G}_{ij}^{\rm F}$,
and  
$
\sum_{i,j=2,3}
{\cal G}_{ij}^{\rm F}$,
respectively.  Here, ${\cal G}_{12}^{\rm F}$ and
${\cal G}_{23}^{\rm F}$ are 
clearly inequivalent because 
they connect  bonds with alternating hopping-pairing amplitudes.
Furthermore, it is seen that
the staggered modulation of the  LDOS  becomes
most conspicuous when scanned along the $a$- or $b$-axis [see the inset of  
Fig.~3(a)],
 because on these bonds 
 the circulating ${\bbox {\cal V}}({\bbox r})$ field becomes parallel to
the bond directions.
From now on, we shall concentrate on the LDOS at the point
${\bbox r}=(0,i_y+1/2)$ with lattice unit.
To evaluate
 ${\cal G}_{ij}^{\rm F}(\ii\omega)$,
we  shall use  the following two approaches which may be complementary to each 
other:
(I) {\it  gradient expansion}, and
(II)  {\it uniform $\bbox{\cal V}$ approximation}.

\noindent
(I) {\it  Gradient expansion}:

First,
we expand Eq.~(\ref{modchiDelta}) with respect to ${\cal V}_{ij}$ up to 
first order
 as
$
\chi_{ij}\sim\chi-
{(-1)^{i_x+j_y}}
\Delta{\cal V}_{ij}
$,
and
$
\Delta_{ij}\sim\Delta+
{(-1)^{i_x+j_y}}
\chi{\cal V}_{ij},
$
which gives
$
\bar U_{ij}^d
=U_{ij}^{d}
+\delta U_{ij}^d
$
with
$
\delta U_{ij}=
{(-1)^{i_x+j_y}}
[\Delta\tau^3
+(-1)^{i_y+j_y}\chi\tau^1]{\cal V}_{ij}.
$
Then, we treat the effect of  $\delta U_{ij}$
within the Born approximation.
 In this case, we can take  account of the circulating  configuration of
 the $\bbox{\cal V}({\bbox r})$ field. 
 
The LDOS on the bonds is written as
$
N({\bbox r}, \omega)/
x \alpha^2
$
$=
\bar N_0(\omega)+\delta \bar N({\bbox r}, \omega).
$
The uniform counterpart is given by 
$
\bar N_0(\omega)
$
$=-{1\over \pi}
\sum_{\bbox k} \cos^2{k_x\over2}
$
${\rm Im} [{\cal G}^{\rm F}_0({\bbox k},\ii\omega)]_{11}\mid_{i\omega\to 
\omega+i\delta},
$
where
$
{\cal G}^{\rm F}_0({\bbox k},\ii\omega)
=
{U_{\bbox k}/(\ii\omega-E_{\bbox k})}+{V_{\bbox k}/(\ii\omega+E_{\bbox k})},
$
$
U_{\bbox k}={1\over2}[1+({\gamma_k\tau^3+\eta_k\tau^1)/ E_{\bbox k}}]
$, 
$
V_{\bbox k}={1\over2}[1-{(\gamma_k\tau^3+\eta_k\tau^1)/ E_{\bbox k}}],
$
and
$
E_{\bbox k}=\sqrt{\gamma_k^2+\eta_k^2}
$
with
$
\gamma_{\bbox k}=-J\chi[
\cos k_x+\cos k_y
+\tilde t_2 \cos k_x\cos k_y
+\tilde t_3(\cos 2 k_x+\cos 2 k_y)]+a_0
$,
and
$
\eta_{\bbox k}=+J\Delta(\cos k_x-\cos k_y)
$.
We took account of the second and third nearest neighbor hopping of the 
fermions
[$\tilde t_2=-0.550$ and $\tilde t_3=0.087$]
to reproduce the
real band structure of 
Bi$_2$Sr$_2$CaCu$_2$O$_{8+\delta}$
as measured by angle-resolved photo emission spectroscopy\cite{Norman98}.

The perturbation term causes the unit cell doubling  and is written in Fourier
 space as
$
\delta H^{\rm F}
=
-
\sum_{{\bbox k}\in \rm RZ}
\sum_{{\bbox q}\sim \rm small}
\sum_\sigma[
\psi_{{\bbox k}+{{\bbox q}\over2}+{\bbox Q}\sigma}^{\dagger}C_{\bbox k}({\bbox 
q})\psi_{{\bbox k}-{{\bbox q}\over2}\sigma}
+{\rm H.c.}].
$
Recalling the expression for  ${\bbox h}({\bbox q})$, we obtain 
$
 C_{\bbox k}({\bbox q})
=\Delta  C^+_{\bbox k}({\bbox q})\tau^3+\chi C^-_{\bbox k}({\bbox q})\tau^1
$,
where
$
C^\pm_{\bbox k}({\bbox q})
=
\pi J  
 {1\over |{\bbox q}|^2}
{\lambda_a |{\bbox q}|\over 1+\lambda_a |{\bbox q}|}
\left[
q_y \sin k_x\pm q_x \sin k_y\right],
$
RZ means the reduced zone $| k_x |+| k_y |\leq \pi$,
and  ${\bbox Q}=(\pi,\pi)$. 
We obtain
\be
&&\delta \bar N({\bbox  r},\omega)
\sim
{(-1)^{i_y}\over 4}
{\cal V}_x({\bbox r})
\sum_{{\bbox k}\in\rm RZ}
\sin^2 k_x\non\\
&&
\left[
L_{\bbox k}^+ \delta(\omega;E_{\bbox k},E_{{\bbox k}+{\bbox Q}}
)
\right.
+
L_{\bbox k}^-
\delta(\omega;-E_{\bbox k},-E_{{\bbox k}+{\bbox Q}})
\non\\
&+&
N_{\bbox k}^+
\delta(\omega;E_{\bbox k},-E_{{\bbox k}+{\bbox Q}})
+\left.
N_{\bbox k}^-\delta(\omega;-E_{\bbox k},E_{{\bbox k}+{\bbox Q}})
\right],
\en
where ${\bbox r}=(0,i_y+1/2)$, 
$
\delta(\omega;x,y)\equiv
[\delta(\omega-x)-\delta(\omega-y)]
/(x-y)
$,
$
L_{\bbox k}^\pm=
\Delta
(1+{\gamma_+}{\gamma_-}-{\eta_+}{\eta_-}\pm{\gamma_+}\pm{\gamma_-})
\pm
\chi ({\eta_+}+{\eta_-}\pm{\gamma_+}{\eta_-}\pm{\eta_+}{\gamma_-})
$,
$
N_{\bbox k}^\pm=
\Delta
(1-{\gamma_+}{\gamma_-}+{\eta_+}{\eta_-}\pm{\gamma_+}\mp{\gamma_-})
\pm
\chi({\eta_+}-{\eta_-}\mp{\gamma_+}{\eta_-}\mp{\eta_+}{\gamma_-})
$,
$
\gamma_+={\gamma_{{\bbox k}+{\bbox Q}}/ E_{{\bbox k}+{\bbox Q}}},
$
$
\gamma_-={\gamma_{\bbox k}/ E_{\bbox k}},
$
$
\eta_+={\eta_{{\bbox k}+{\bbox Q}}/ E_{{\bbox k}+{\bbox Q}}},
$
and
$
\eta_-={\eta_{\bbox k}/ E_{\bbox k}}.
$
Because of the staggerdness,  a fermion 
with $\bbox k$ in the reduced zone is scattered  
to ${\bbox k}+{\bbox Q}$ in the second zone. 
 
 In Fig.~3(a), we show the profile of  
 $N({\bbox  r},\omega)/x \alpha^2$ 
 at the points  A$(0,i_y+1/2)$, B$(1/2,i_y+1)$, C$(0,i_y+3/2)$, and 
D$(-1/2,i_y+1)$
 by choosing $i_y$ to 
satisfy ${\cal V}_x({\bbox{r}})=0.1$. 
As a rough estimate, this corresponds to a lattice position of 
$i_y=4$ if $\lambda_a=10$.
 We used
 $a_0=0.05 \chi J$ and
 $\Delta/\chi=0.2$. 
 Numerical integration was performed by dividing the Brillouin zone into a
$320\times 320$ mesh.
 Note that at B and D, 
 $\delta \bar N({\bbox  r},\omega)$
 almost vanishes and the LDOS is just given by 
 $\bar N_0(\omega)$,
because  ${\bbox{\cal V}}({\bbox{r}})$ becomes almost perpendicular to these 
bond directions.
 The modulation pattern at the other points can  be  read off from Fig.~2.

  It is remarkable that,
 inside the overall V-shaped profile with the sharp peaks at
$\tilde\omega\equiv\omega/\chi J=\pm 0.323$ associated with the $d$-wave 
superconducting
gap, there appears additional staggered structure 
 around $\tilde\omega=\pm 0.179$ and $\tilde\omega=0.226$.
 This  structure comes from 
 resonant scattering between the fermions with
${\bbox k}$ and ${\bbox k}+{\bbox Q}$.
As $\omega$ increases from zero, the energy contours 
$E_{{\bbox k}}=\omega$ and $E_{{\bbox k}+{\bbox Q}}=\omega$ first touch
on the reduced zone boundary at $\tilde\omega=\pm 0.179$, 
as indicated in Fig.~3(b), and resonance occurs. 
Then, at $\tilde\omega=\pm 0.226$, they touch 
again at $(\pi/2,\pi/2)$ and the second resonance occurs. 
The second resonance comes up only in the electron ($\omega>0$) side due to 
the matrix element effect [$L_{\bbox k}^-$ vanishes at $(\pi/2,\pi/2)$].
It is naturally expected  that this resonant scattering may open
a gap  in the fermion excitation spectrum
if we go beyond the perturbative scheme.
This point can  be confirmed through the exact treatment under
the uniform $\bbox{\cal V}$ approximation shown below.

\noindent
(II) {\it Uniform $\bbox{\cal V}$ approximation}:

Next we consider the case of
uniform ${\bbox {\cal  V}}=({\cal V}_x,{\cal V}_y)$ which may locally
capture the effects of  the circulating ${\bbox {\cal V}}( {\bbox r})$.
In this case,  we  exactly diagonalize 
the fermion Hamiltonian.
The LDOS on the bonds is written in a form
$
N(\omega)/
$
$x \alpha^2=
\tilde N_0(\omega)\pm\delta \tilde N(\omega),
$
where $\pm$ signs alternate from
bond to bond.
As was inferred  from the perturbative analysis, 
the unit cell doubling 
splits
 the one-particle spectrum into two branches [see Fig.~3(e)], 
$
\pm E_{\bbox k} ^+
$ 
and
$
\pm E_{\bbox k} ^-
$.

\begin{figure}
\epsfxsize=3.5in
\centerline{\epsffile{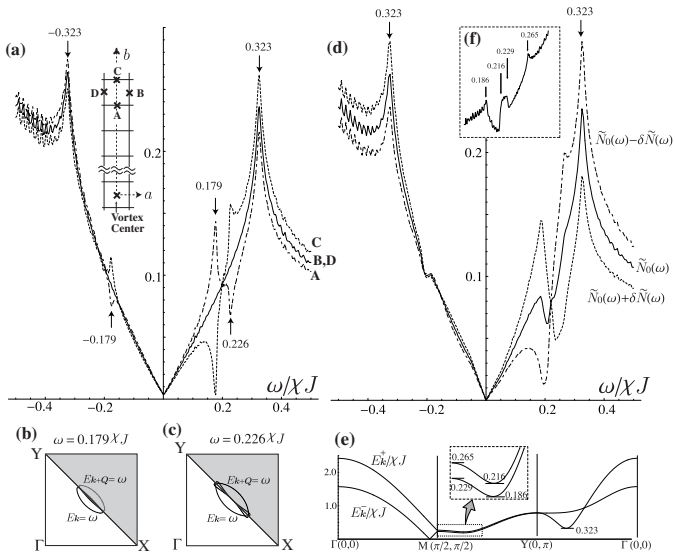}}
\smallskip
\caption{
(a) LDOS profile at the points
A, B, C, and D indicated in the inset.
The LDOS at B and C are just $\bar N_0(\omega)$.
The energy contours $E_{{\bbox{k}}}=\omega$ and
$E_{{\bbox{k}}+{\bbox Q}}=\omega$ touch at
 $\tilde\omega=0.179$ 
and 
$\tilde\omega=0.226$, as indicated in (b) and (c), respectively.
The sharp peaks at $\tilde\omega=\pm0.323$ are associated with
 the superconducting gap. 
The small wiggles outside the V-shaped profile come from numerical
fluctuations. 
(d) Profile 
of $\tilde N_0(\omega)$, and
$\tilde N_0(\omega)\pm\delta \tilde N(\omega)$
for the uniform gauge field ${\bbox {\cal V}}=(-0.1,0)$.
(e) The band dispersion of 
the splitted bands $E_{\bbox k}^\pm$
along the path $\Gamma(0,0)\to{\rm M}(\pi/2,\pi/2)\to
{\rm Y}(0,\pi)\to\Gamma$. Fine
band splittings
on the reduced zone boundary are magnified
 in the inset.
(f) Fine structure 
of $\tilde N_0(\omega)$ around $\tilde\omega\sim 0.2$,
detected with higher numerical resolution.
}
\end{figure}

 In Fig.~3(d), we show the profile of  
 $\tilde N_0(\omega)$ 
 and $\tilde N_0(\omega)\pm\delta \tilde N(\omega)$ for 
 ${\bbox {\cal  V}}=(-0.1,0)$, the
 direction and strength of which 
correspond to ${\bbox {\cal  V}}({\bbox r})$
around the points B, D, and A,C  in Fig.~3(a), respectively.
 We  used the same parameter set as in the case of Fig.~3(a).
We see 
 that the staggerd modulation profile, $\tilde N_0(\omega)\pm\delta \tilde 
N(\omega)$,
 is in remarkable agreement 
 with that obtained by the perturbative analysis.
The structures correspond to  van-Hove singularities associated with
the gap opening on the reduced zone boundary
as shown in Fig.~3(e).
However, a striking difference is that the dip structure around 
$\tilde\omega=0.2$
appears even in the uniform counterpart $\tilde N_0(\omega)$.
This suggests that in reality
 the dip structure may be detected not only on the bonds but also at sites.
In Fig.~3(d), 
due to
numerical resolution [320$\times$320 mesh of the Brillouin zone]
we resolve structures around
$\tilde\omega=\pm 0.2$ and $\tilde\omega=0.26$.
As indicated in Fig.~3(e), there are fine
 band splittings on the reduced zone boundary  
leading to van Hove singularities at
$\tilde\omega=0.186$, 0.216, 0.229, and 0.265.
The corresponding fine structure in $\tilde N_0(\omega)$ 
could be detected with much higher
numerical resolution [720$\times$720 mesh of the Brillouin zone],
as shown in Fig.~3.(f).

We note that, in both approximations,
the modulated structure in the LDOS is predominant on the particle side 
($\omega>0$).
We can understand this asymmetry by first turning off the superconductivity 
and considering the effect of
unit cell doubling.   Since we are doping with holes, the gaps being opened by 
the unit cell doubling are on the empty side of the Fermi surface. Matrix 
element effects preserve this particle-hole asymmetry even after we turn on
the superconductivity.

Combining the results obtained through the gradient expansion and the
uniform $\bbox{\cal V}$ approximation, we may reasonably say that 
 the signature of the unit cell doubling may be most prominently detected
through the characteristic dip structure inside the V-shaped profile.
The structure predicted here is very specific and its observation will be
a strong confirmation of the SU(2) vortex model.

J.K. is supported by a Monbusho Grant for overseas research.
P.A.L. and X.G.W. acknowledge support by NSF under the MRSEL
Program DMR 98-08491.
X.G.W. also acknowledges support by NSF grant No. DMR 97-14198.

\end{document}